%% file: main.tex
\documentclass[letterpaper, 10 pt, conference]{ieeeconf}  

\IEEEoverridecommandlockouts                       
\usepackage{cite}
\usepackage{graphicx}

\usepackage{amsmath, amssymb, amsfonts} 
\usepackage{booktabs} 
\usepackage{mathtools}
\usepackage{commath}

\usepackage{microtype}
\usepackage{mathdots}
\usepackage{mathtools}
\usepackage{bm}

\usepackage{eucal}

\usepackage{tikz}
\usetikzlibrary{decorations.markings,arrows.meta} 
\usepackage{pgfplots}
\pgfplotsset{every axis/.append style={line width=1pt}}
\usetikzlibrary{shapes, fit, decorations.markings, matrix, plotmarks, positioning, fit, spy, patterns, shadows, calc, backgrounds, arrows.meta}
\pgfplotsset{compat=1.6}
\tikzset{
  trajectory path/.style={
    line width=0.65pt,
    postaction={decorate},
    decoration={
      markings,
      mark=at position 0.40 with {
        \arrow{Latex[length=1.1mm, width=0.7mm]}
      }
    }
  }
}

\pgfplotsset{every tick label/.append style={font=\tiny}}
\pgfplotsset{every axis/.append style={
                    label style={font=\footnotesize}
                    }}

\usepackage{algorithm} 
\usepackage{algpseudocode} 

\usepackage[font=footnotesize,caption=false]{subfig}
\usepackage{graphicx}
\usepackage[abs]{overpic}

\usepackage{inputenc}
\usepackage[nolist]{acronym}
\usepackage{siunitx}
\usepackage{booktabs}
\usepackage{lipsum}

\usepackage{enumerate}
\newcounter{subeqn} %

\newtheorem{theorem}{Theorem}
\newtheorem{corollary}{Corollary}
\newtheorem{lemma}[theorem]{Lemma}
\newtheorem{proposition}{Proposition}

\newtheorem{definition}{Definition}

\newtheorem{remark}{Remark}
\newtheorem{problem}{Problem}

\newcommand{\ts}[1]{{\textnormal{#1}}}
\newcommand{\ie}{\emph{i.e.,}\ }

\newcommand{\mc}{\mathcal}

\newcommand{\tb}{\textbf}
\newcommand{\mbb}{\mathbb}

\newcommand{\Rset}{\mathbb{R}}

\newcommand{\TxM}[2]{{\mathrm{T}_{#1}{#2}}}

\newif\ifproves
\provestrue

\allowdisplaybreaks

\begin{acronym}
  \acro{MPC}{Model Predictive Control}
  \acro{PE}{Persistently Exciting}	
  \acro{LMI}{Linear Matrix Inequalities}
  \acro{CBF}{Control Barrier Function}
\end{acronym}
\usepackage{cite}
\title{Control Invariance on Manifolds through Tangent Bundle Splitting}
\author{Luke Georgiou$^{1}$, Ross Drummond$^1$, Pablo R. Baldivieso Monasterios$^{1}$\thanks{$^{1}$ Luke Georgiou, Ross Drummond, and Pablo. R. Baldivieso Monasterios are with the School of Electric and Electronic Engineering, University of Sheffield, Sheffield, UK {\tt\footnotesize \{lgeorgiou3,ross.drummond,p.baldivieso\}@shef.ac.uk}}
}

\begin{document}
\bstctlcite{nourl}
\bstctlcite{nodoi}

\maketitle

\begin{abstract}      
Control invariance for interconnected non-linear systems on manifolds requires conditions under which subsystem safety constraints can be combined to form an invariant set. This paper develops local product type control invariant sets on manifolds using tangent bundle splitting, adapted product coordinates, and subsystem tangency conditions satisfied simultaneously by one admissible input at each state. The resulting conditions establish control invariance of product sets without requiring independent subsystem dynamics, with separate control choices possible when inputs are independently constrained and each acts only on its subsystem. This provides a way to verify subsystem safety without discarding dynamic coupling and is illustrated by a torus example in which adapted coordinates give a curved constraint region a product representation.
\end{abstract}
\section{Introduction}
\label{sec:introduction}
The state of a network of $N$ interconnected systems can be described in Euclidean space by the partition $x=(x_1,\cdots,x_N)\in\Rset^n=\mbb{R}^{n_1}\times\cdots\times\mathbb{R}^{n_N}$, with each $x_i$ being the state of the $i^\ts{th}$ subsystem. In this paper, we assume each subsystem $i$ can be controlled independently, using $u_i\colon\Rset\to\mbb{U}_i$ with $\mbb{U}_i\subset\Rset^{m_i}$ compact, and the dynamics evolve by
\begin{equation*}
    \dot x_i =F\left(x_i,x_{-i},u_i\right),\quad  x_i\in \Rset^{n_i},\quad u_i\in\mbb{U}_i,
\end{equation*}
where $x_{-i}$  collects the states of the other subsystems on $i$. Given this partition, a natural question is whether there exist sets $\Omega_i$ such that an admissible control keeps $x_i\left(t\right)\in\Omega_i$ for every subsystem $i$ and all $t\geq0$, whenever $x_i\left(0\right)\in\Omega_i$ for all $i$. The main tools for answering this question lie in set invariance theory \cite{Blanchini1999} and viability theory \cite{aubin2009viability}. For linear systems, \cite{Hernandez2016} gives conditions for the existence of such sets based on the coupling strength, \ie $\{\Omega_i(\norm{x_{-i}})\}_{i=1}^N$, and \cite{Trodden2026} shows the maximum coupling that allows these sets to exist. These ideas were exploited in distributed control for linear networks in \cite{Baldivieso-Monasterios2021} and \cite{Masero2023} by allowing a time-varying state partition. As a result, partitioning the state is key to obtaining distributed guarantees from a network.

 Set invariance is closely linked to safety, as trajectories remain ``trapped'' within the set. For example, consider a scalar function, known as a \ac{CBF}, $h\colon \Rset^n\to \Rset$; its zero superlevel set $\Omega = \{x\colon h(x) \geq 0\}$ serves as the safety certificate. The authors in \cite{Ames2016} and \cite{Rabiee2025} employ quadratic programmes and Lie derivatives to ensure this level set encloses the system trajectories. The stochastic setting was handled by \cite{Prajna2007}, and probabilistic results were obtained in \cite{WANG2025112393}. An alternative view of safety certification can be achieved via reachability analysis using the Hamilton-Jacobi-Bellman equation \cite{Bansal2018}. For networks, existing work, see \cite{WANG2025112393} and  \cite{Nilsson2016}, examines how subsystem constraints can be combined into an invariant set while accounting for dynamic coupling. Recent work in \cite{Pirastehzad2026} proposes global certification via only local certificates. This implies that the certificates have a product-like structure.

 Geometric control describes dynamics intrinsically on manifolds \cite{Jurdjevic1996, Bullo2005}, allowing system structure to be expressed independently of coordinates.Safety filters based on \ac{CBF}s on manifolds have been explored using backstepping techniques \cite{deSa2025geometric}. Ibuki et al. \cite{Ibuki2020} formulated \ac{CBF}s on $\mbb{S}^2$ for collision avoidance. Networks evolving on manifolds have also been an active research area, as shown in power converters \cite{Doerfler2012} and attitude tracking \cite{lee2011}. For interconnected systems on manifolds, a global state partition may not exist, coordinates are no longer globally defined, and identifying coordinates and subsystems is no longer straightforward. In recent work, Abdelgalil et al. \cite{Abdelgalil2026} show how to model multi-agent systems on a manifold by splitting the tangent bundle via distributions. This type of splitting identifies subsystem velocity directions, but does not by itself provide compatible coordinates or a safety guarantee. In this paper, we construct invariant sets with a local product structure for interconnected systems evolving on manifolds. Following \cite{Abdelgalil2026}, subsystem directions are represented by distributions $\{\mc{D}_i\}_i^{N}$ that are direct sums of the tangent bundle of the ambient manifold $M$. These distributions generalise coordinates in curved spaces and can be assigned to each subsystem; we show that the distributions we build are involutive. This does not, by itself, provide compatible subsystem state coordinates. On these distributions, we introduce the notion of ``adapted chart'' $\Phi\colon M\to\Rset^n$ which satisfies $\Omega=\Phi^{-1}\left(\Omega_1\times\cdots\times\Omega_N\right)$.  As such, each subsystem distribution corresponds to its own coordinate block and, within these coordinates, one admissible input must satisfy all subsystem tangency conditions simultaneously. Furthermore, we give conditions for which a splitting supports local product representations of constraints and allows their control invariance to be verified through subsystem conditions.

 For the problem formulated in Section~\ref{sec:problem-formulation}, this paper:
 \begin{enumerate}[i)]
 \item The paper establishes necessary and sufficient conditions for the existence of coordinates adapted to the given subsystem distributions in Proposition~\ref{prop:commuting-frames}, and Section~\ref{sec:tangent-bundle-splitting}. These coordinates allow subsystem constraints to be represented as a product without assuming the manifold is, globally, a Cartesian product. 
 \item Establishes the control invariance of compact product sets through subsystem tangency conditions satisfied by a common input in Theorem~\ref{thm:local-product-invariance}, and describes the set and conditions intrinsically on $M$ in Section~\ref{sec:tangent_cones}.
 \item Separates feasible subsystem input choices under Cartesian product input constraints and input alignment, without requiring independent subsystem dynamics in Corollary~\ref{cor:blockwise-feasibility}.
 \end{enumerate}
 The existence of adapted charts and their associated distributions are conditions guaranteeing the existence of an agent interpretation of the non-linear dynamics and pave the way for decentralised and distributed safety certificates.

\textit{\underline{Notation}:} We use standard geometric notation from
\cite{doCarmo1992,Lee2018}. Let $M$ be a smooth $n$-dimensional manifold without boundary,
with tangent spaces $\TxM{x}{M}$ and tangent bundle $TM=\bigsqcup_{x\in M}\TxM{x}{M}$. The
differential of a smooth map $H:M\to N$ at $x$ is $dH_x:\TxM{x}{M}\to\TxM{H(x)}{N}$, and
$\Gamma(\TxM{}{M})$ is the set of smooth vector fields. For $X,Y\in\Gamma(\TxM{}{M})$, the
Lie bracket is $[X,Y]h=X(Yh)-Y(Xh)$ for $h\in C^\infty(M)$, where $Xh:=dh(X)$. A smooth
distribution $\mc{D}\subset TM$ of constant rank $r$ is a smooth vector subbundle whose
fibres $\mc{D}(x)\subseteq\TxM{x}{M}$ are $r$-dimensional. It is \emph{involutive} if its
smooth local sections are closed under Lie brackets, and \emph{integrable} if every point
lies in an immersed submanifold $L$ with $\TxM{y}{L}=\mc{D}(y)$ for all $y\in L$. By the
Frobenius theorem \cite{Lee2013}, $\mc{D}$ is integrable if and only if it is involutive, and
then every point has a neighbourhood $W$ and a submersion $\zeta:W\to\Rset^{n-r}$ with
$\ker d\zeta=\mc{D}|_W$. The symbol $\oplus$ denotes the internal direct sum of subspaces, and
$\operatorname{co}\mc{A}$ the convex hull of a set $\mc{A}$. For $K\subset M$, $x\in K$ and a chart $(O,\varphi)$ at
$x$, the \emph{contingent cone} of $K$ at $x$, adapted from \cite{Rockafellar1998Variational},
is
\begin{multline*}
T_K(x):=\bigl\{v\in\TxM{x}{M}:\exists\,x_k\in K\cap O,\ h_k\downarrow0\ \text{with}\\
h_k^{-1}\bigl(\varphi(x_k)-\varphi(x)\bigr)\to d\varphi_x(v)\bigr\}.
\end{multline*}
The same argument gives $dH_x(T_K(x))=T_{H(K)}(H(x))$ for a diffeomorphism $H$ between open sets containing $K$, and $T_K(x)=\TxM{x}{M}$ for $x$ in the interior of $K$. For a control system $\dot x=F(x,u)$ on $M$ with input set
$\mbb{U}\subset\Rset^m$, an input is a measurable $\mu:[0,\infty)\to\mbb{U}$, and
$\psi(\cdot,x_0,\mu)$ denotes the corresponding Carathéodory solution with
$\psi(0,x_0,\mu)=x_0$. A set $\Omega\subset M$ is \emph{control invariant} if for every
$x_0\in\Omega$ there is an input $\mu$ such that $\psi(t,x_0,\mu)$ is defined and lies in
$\Omega$ for all $t\geq0$. We write $\mc{F}(x)=\{F(x,u):u\in\mbb{U}\}\subset\TxM{x}{M}$. The following lemma is the only viability result used in the paper.
\begin{lemma}[Viability in a chart]
\label{lem:viability}
Let $F(x,u)=f(x)+g(x)u$ with $f$ and $g$ smooth, and let $\mbb{U}\subset\Rset^m$ be
nonempty, convex and compact. Let $(O,\varphi)$ be a chart and $K\subset O$ nonempty and
compact. If
\begin{equation}
\mc{F}(x)\cap\operatorname{co}T_K(x)\neq\varnothing\quad\text{for all } x\in K,
\label{eq:controlled-nagumo}
\end{equation}
then $K$ is control invariant.
\end{lemma}
 
\section{Problem Formulation:}
\label{sec:problem-formulation}
Given the interconnected dynamics on a manifold
\begin{equation}
  \dot{x} = F(x,u) = f(x) + g(x)u,\quad u\in\mbb{U}\subset\Rset^m,
  \label{eq:net_dyn}
\end{equation}
where $x\in M$, $f\in\Gamma(\TxM{}{M})$, $g(x)=[g_1(x)\ \cdots\ g_m(x)]$ with
$g_j\in\Gamma(\TxM{}{M})$, and $\mbb{U}$ is nonempty, convex and compact. Subsystems are not blocks of the state vector since $M$ may not be a product and coordinates may not respect interconnection. We first formally define our notion of tangent bundle splitting.
\begin{definition}[Complementary tangent-bundle splitting]
\label{def:complementary-splitting}
For smooth, constant rank $n_i$, distributions $\mc{D}_i \subset TM$, $i \in \mc{I} = \{1,\ldots,N\}$ define a \emph{complementary tangent bundle splitting} if $\TxM{x}{M}=\bigoplus_{i\in\mc{I}}\mc{D}_i(x)$ for every $x\in M$.
\end{definition}
Each $v_x\in\TxM{x}{M}$ has a unique decomposition $v_x=\sum_{i\in\mc{I}}v_i$ with $v_i\in\mc{D}_i(x)$ and $\sum_{i\in\mc{I}}n_i = n$. The splitting fixes subsystem directions without reference to coordinates, while the dynamics and the inputs in \eqref{eq:net_dyn} may still couple the subsystems. On $\Rset^n= \prod_{i\in\mc{I}}\Rset^{n_i}$, let $\jmath_i:\Rset^{n_i}\to\Rset^n$ be the injection onto the $i$-th block,
$E_i:=\jmath_i(\Rset^{n_i})$ the $i$-th coordinate subspace,
$P_i:=\operatorname{blkdiag}(0_{n_1},\dots,\mbb{I}_{n_i},\dots,0_{n_N})$ the projection onto $E_i$
along the other blocks, and $e_{i,a}$, $a=1,\dots,n_i$, the standard basis of $E_i$. To impose constraints subsystem by subsystem, we need coordinates
$z=(z_1,\dots,z_N)$, $z_i\in\Rset^{n_i}$, in which motion along $\mc{D}_i$ changes $z_i$
only, that is, a diffeomorphism
\begin{equation}
\begin{gathered}
\Phi:U\to V_1\times\cdots\times V_N,\\
d\Phi_x(\mc{D}_i(x))=E_i,\quad x\in U,\ i\in\mc{I},
\end{gathered}
\label{eq:candidate-product-chart}
\end{equation}
where $U\subset M$ and $V_i\subseteq\Rset^{n_i}$ are open and $T_{\Phi(x)}(V_1\times\cdots\times V_N)$ is identified with $\Rset^n$. In such a chart,
subsystem constraint sets $\Omega_i\subset V_i$, each nonempty and compact, combine into the candidate set 
\begin{equation}
\Omega:=\Phi^{-1}\Bigl(\prod_{i\in\mc{I}}\Omega_i\Bigr)\subset U,
\label{eq:candidate-set}
\end{equation}
which is compact and contained in the chart domain. The construction is local and does not
assume that $M$ is globally a product. Our objective, shown in Figure~\ref{fig:adapted-product-chart} is to determine when the given splitting supports a local product representation of the constraints and when the resulting set is control invariant. 
\begin{figure*}[t!]
  \centering
  \input{adaptedproductchart}
    \caption{ Adapted product chart for $N=2$. Thin curves are integral curves of $\mc{D}_1$
  (blue) and $\mc{D}_2$ (teal); $\Phi$ maps them to coordinate lines, so
  $d\Phi_x(\mc{D}_i(x))=E_i$. The product set $\Omega_1\times\Omega_2$ pulls back to
  $\Omega=\Phi^{-1}(\Omega_1\times\Omega_2)$, whose faces $\{z_1\in\partial\Omega_1\}$ are
  integral curves of $\mc{D}_1^c=\mc{D}_2$ and whose faces $\{z_2\in\partial\Omega_2\}$ are
  integral curves of $\mc{D}_1$. On the face $\Sigma_1$, the component $F_2$ is tangent to
  the face, so only $F_1$, equivalently $\widehat F_1$, determines whether the trajectory
  leaves $\Omega$.}
    \label{fig:adapted-product-chart}
\end{figure*}
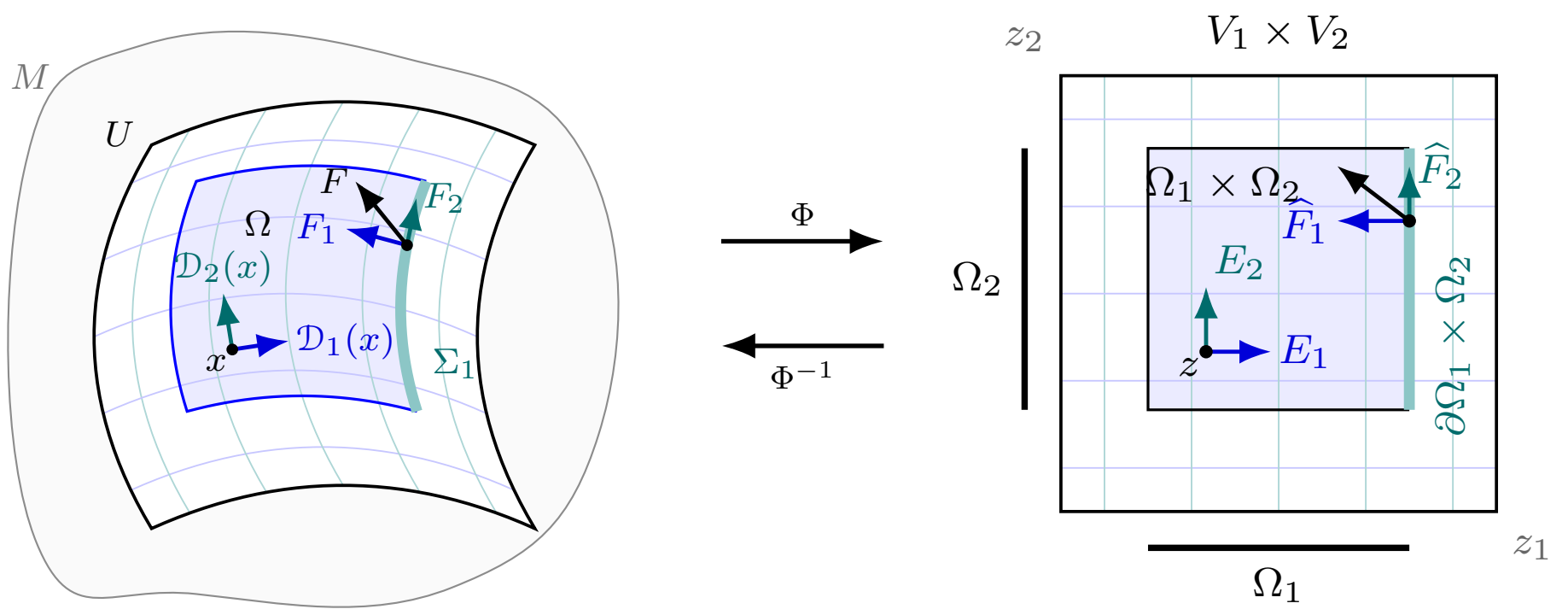
\begin{problem}
\label{prob:subsystem safety}
Given the network \eqref{eq:net_dyn}, and the splitting in Definition~\ref{def:complementary-splitting},
\begin{enumerate}
\item determine when charts of the form \eqref{eq:candidate-product-chart} exist;
\item give conditions on the subsystem velocities, imposed through the sets $\Omega_i$ and
met by a single admissible input at each $x\in\Omega$, under which \eqref{eq:candidate-set} is control invariant;
\item describe the resulting invariant set and the invariance condition on $M$,
independently of the chart, so that the safety guarantee does not depend on the coordinates used.
\end{enumerate}
\end{problem}

\section{Control invariance from tangent-bundle splitting}
\label{sec:tangent-bundle-splitting}

For the network \eqref{eq:net_dyn} and the complementary splitting introduced in Definition~\ref{def:complementary-splitting}, we first describe the associated subsystem projections and conditions under which the distributions admit compatible local product coordinates. These coordinates decompose the state only under an integrability condition characterised in Section~\ref{sec:charts}.

\subsection{Compatible subsystem structure}
\label{sec:velocities}
The first step, smoothly assigning each tangent vector its subsystem components, requires no integrability.
\begin{lemma}[Smooth subsystem projections]\label{lem:subsystem-projections}
  Let $TM=\bigoplus_{i\in\mc{I}}\mc{D}_i$ be a complementary splitting  with inclusions $\iota_i:\mathcal{D}_i\hookrightarrow TM$. For $i\in\mc{I}$, let $\Pi_i:TM\to\mc{D}_i$ restrict on each $\TxM{x}{M}$ to the projection onto $\mc{D}_i(x)$ along $\mc{D}_i^{c}(x)$. Then each $\Pi_i$ is a smooth bundle homomorphism over $\mathrm{id}_M$, and 
\begin{equation*}
  \Pi_i\circ\iota_i=\mathrm{id}_{\mc{D}_i},\quad
  \Pi_i\circ\iota_j=0\ (i\neq j),\quad
  \sum_{i\in\mc{I}}\iota_i\circ\Pi_i=\mathrm{id}_{TM}.
\end{equation*}
In particular, $\hat{\Pi}_i:=\iota_i\circ\Pi_i$ satisfies $\hat{\Pi}_i\circ\hat{\Pi}_j=\delta_{ij}\hat{\Pi}_i$.
\end{lemma}

\begin{proof}
The identities hold fibrewise by uniqueness of the decomposition $v=\sum_{j\in\mc{I}}v_j$,
$v_j\in\mc{D}_j(x)$, and
$\hat{\Pi}_i\circ\hat{\Pi}_j=\iota_i\circ(\Pi_i\circ\iota_j)\circ\Pi_j=\delta_{ij}\hat{\Pi}_i$. Smoothness is local. Around any point of $M$, fix a chart $(W,\varphi)$ on which each
$\mc{D}_i$ has a smooth local frame $(X^i_1,\dots,X^i_{n_i})$, and let
$A:W\to\Rset^{n\times n}$ have as columns the coordinate representations of
$X^1_1,\dots,X^1_{n_1},\dots,X^N_1,\dots,X^N_{n_N}$. Then $A$ is smooth, and $A(x)$ is
invertible for every $x\in W$ by the direct-sum hypothesis. In the induced chart of $TW$,
$\hat{\Pi}_i$ reads $(x,v)\mapsto\bigl(x,A(x)P_iA(x)^{-1}v\bigr)$, which is smooth because
$A^{-1}=\operatorname{adj}(A)/\det A$. Hence $\hat{\Pi}_i$ is smooth, and so is $\Pi_i$,
since $\mc{D}_i$ is an embedded submanifold of $TM$ containing $\hat{\Pi}_i(TM)$.
\end{proof}

Applying these projections gives the subsystem velocities
\begin{equation}
    \begin{split}
F_i(x,u)&=\Pi_i(x)F(x,u)\in\mc{D}_i(x),\\
F(x,u) &=\sum_{i\in\mc{I}}F_i(x,u).
    \end{split}
\label{eq:velocity-components}    
\end{equation}
This decomposition is chart-free and exists for every splitting.
\subsection{Adapted product charts}
\label{sec:charts}

Interpreting the distributions as local subsystem state spaces requires additional integrability and compatibility.  A state of subsystem $i$ is a submersion
$z_i$ from an open subset of $M$ onto an open subset of $\Rset^{n_i}$ whose rate depends only on the $i$-th component. For every tangent vector $v$ over its domain,
\begin{equation}
dz_i(v)=dz_i\bigl(\hat{\Pi}_iv\bigr).
\label{eq:subsystem-state}
\end{equation}
By a rank count, \eqref{eq:subsystem-state} holds iff
$\ker dz_i=\mc{D}_i^c$. Collecting states of all subsystems gives a chart of the form
\eqref{eq:candidate-product-chart}.

\begin{definition}[Locally product-compatible splitting]
\label{def:product-compatible-splitting}
A diffeomorphism $\Phi=(z_1,\dots,z_N):U\to V_1\times\cdots\times V_N$ satisfying
\eqref{eq:candidate-product-chart} is an \emph{adapted product chart}. A complementary
splitting is \emph{locally product compatible} if every point of $M$ has an open
neighbourhood that is the domain of an adapted product chart.
\end{definition}
For a diffeomorphism $\Phi=(z_1,\dots,z_N)$, condition \eqref{eq:candidate-product-chart} is equivalent to $\ker dz_i=\mc{D}_i^c$ for every $i$, that is, to each $z_i$ being a state of subsystem $i$ in the sense of \eqref{eq:subsystem-state}. The coordinate $z_i$ is constant along $\mc{D}_i^c$: motion in the directions of the other subsystems does not change it (see Figure~\ref{fig:adapted-product-chart}). The next result characterises when such charts exist.
\begin{proposition}[Product-compatible splittings]
\label{prop:commuting-frames} 
Let $TM=\bigoplus_{i\in\mc{I}}\mc{D}_i$ be a smooth complementary splitting,
$n_i:=\operatorname{rank}\mc{D}_i$, and
$\mc{D}_i^{c}:=\bigoplus_{j\in\mc{I}\setminus\{i\}}\mc{D}_j$.
The following are equivalent:
\begin{enumerate}[i)]
\item the splitting is locally product compatible,
\item $\mc{D}_i^{c}$ is involutive for every $i\in\mc{I}$,
\item around every point of $M$ there are local frames $\{X_{i,a}\}_{a=1}^{n_i}$ of $\mc{D}_i$ such that $[X_{i,a},X_{j,b}]\in\Gamma(\mc{D}_i\oplus\mc{D}_j)$ for all
$i,j\in\mc{I}$ and all $a,b$,
\item around every point of $M$ there are local frames $\{X_{i,a}\}_{a=1}^{n_i}$ of
$\mc{D}_i$ such that $[X_{i,a},X_{j,b}]=0$ for all $i,j\in\mc{I}$ and all $a,b$.
Moreover, the condition in $iii)$ holds for one choice of local frames iff it
holds for every choice.
\end{enumerate}
\end{proposition}
\begin{proof}
$i)\Rightarrow iv)$: take the coordinate fields $X_{i,a}:=d\Phi^{-1}(e_{i,a})$ of an adapted
chart. $iv)\Rightarrow iii)$: immediate. $iii)\Rightarrow ii)$: for $X=\sum_a f^aX_{i,a}$ and $Y=\sum_b g^bX_{j,b}$, their bracket is $[X,Y]= \sum_{a,b}\bigl(f^ag^b[X_{i,a},X_{j,b}]  +f^aX_{i,a}(g^b)X_{j,b} - g^bX_{j,b}(f^a)X_{i,a}\bigr)\in\Gamma(\mc{D}_i\oplus\mc{D}_j)$, proving frame independence. By Lemma~\ref{lem:subsystem-projections},
every local section of $\mc{D}_k^{c}$ is a sum of local sections of $\mc{D}_j$,
$j\neq k$, and $\mc{D}_i+\mc{D}_j\subseteq\mc{D}_k^{c}$ for $i,j\neq k$.
Finally, $ii)\Rightarrow i)$: fix $x^{\ast}\in M$. By the Frobenius theorem, for each $i$ there is a submersion $z_i$ from a neighbourhood of $x^{\ast}$ into $\Rset^{n_i}$ with
$\ker dz_i=\mc{D}_i^{c}$. Let $\Phi:=(z_1,\dots,z_N)$ on the intersection of these
neighbourhoods. For $j\neq i$, $\mc{D}_i\subseteq\mc{D}_j^{c}$ gives
$dz_j(\mc{D}_i)=\{0\}$, while $\ker dz_i\cap\mc{D}_i=\{0\}$ makes
$dz_i|_{\mc{D}_i}$ injective, hence bijective onto $\Rset^{n_i}$. Thus
$d\Phi_x(\mc{D}_i(x))=E_i$ for all $i$, and $d\Phi_x$ is invertible. By the inverse
function theorem, $\Phi$ maps an open $W\ni x^{\ast}$ diffeomorphically onto an open set;
choose open $V_i$ with $\Phi(x^{\ast})\in V_1\times\cdots\times V_N\subseteq\Phi(W)$ and set $U= W\cap\Phi^{-1}(V_1\times\cdots\times V_N)$.
\end{proof}

\begin{remark}
\label{rem:pairwise}
For $N=2$, condition $ii)$ states that $\mc{D}_1$ and $\mc{D}_2$ are involutive. For
$N\geq3$, involutivity of every $\mc{D}_i$ is necessary but not sufficient. On $\Rset^3$,
the line fields $\mc{D}_1=\operatorname{span}\{\partial_x\}$,
$\mc{D}_2=\operatorname{span}\{\partial_y+x\partial_z\}$ and
$\mc{D}_3=\operatorname{span}\{\partial_z\}$ are complementary and involutive, but
$[\partial_x,\partial_y+x\partial_z]=\partial_z\notin\mc{D}_1\oplus\mc{D}_2=\mc{D}_3^c$.
\end{remark}

In adapted coordinates, the projections of Lemma~\ref{lem:subsystem-projections} become
constant coordinate projections, so the velocity decomposition
\eqref{eq:velocity-components} becomes a decomposition into coordinate blocks.
\begin{lemma}[Coordinate form of projections]
\label{lem:coordinate-projections}
Let $\Phi=(z_1,\dots,z_N)\colon U \to V_1\times\cdots\times V_N$ be an adapted product chart, with $z_i=(z_{i,1},\dots,z_{i,n_i})$.  Then, for $x\in U$, $d\Phi_x\circ\hat{\Pi}_i(x)=  P_i\circ d\Phi_x$, or equivalently
\begin{equation}
    \hat{\Pi}_i= \sum_{a=1}^{n_i}\frac{\partial}{\partial z_{i,a}}\otimes dz_{i,a}.
\label{eq:coordinate-projection-identity}
\end{equation}
Conversely, a diffeomorphism $\Phi:U\to V_1\times\cdots\times V_N$ satisfying the second
identity for all $i\in\mc{I}$ is an adapted product chart.
\end{lemma}
\begin{proof}
Fix $x\in U$ and $v\in\TxM{x}{M}$. Adaptedness gives $\partial/\partial z_{j,b}|_x=(d\Phi_x)^{-1}e_{j,b}\in\mc{D}_j(x)$, so $v=\sum_{j\in\mc{I}}\sum_{b=1}^{n_j}dz_{j,b}(v)\,\partial/\partial z_{j,b}|_x$ is the
decomposition of $v$ along the splitting. By its uniqueness, $\hat{\Pi}_i(x)v=\sum_{a=1}^{n_i}dz_{i,a}(v)\,\partial/\partial z_{i,a}|_x$, and applying
$d\Phi_x$ gives the first identity. Conversely, taking images in the second identity gives $d\Phi_x(\mc{D}_i(x))=P_i\,d\Phi_x(\TxM{x}{M})=E_i$.
\end{proof}

\section{Control invariance of product sets}
\label{sec:tangent_cones}
Let $\Phi\colon U\to \prod_{i\in\mc{I}}V_i$ be an adapted product chart and consider the candidate set \eqref{eq:candidate-set}. Since $\Omega\subset U$, the transformation rule for contingent tangent cones gives
\begin{equation}
d\Phi_x\bigl(T_{\Omega}(x)\bigr)=T_{\Phi(\Omega)}(\Phi(x)),\qquad x\in\Omega,
\label{eq:cone-transformation}
\end{equation}
with $\Phi(\Omega)=\prod_{i\in\mc{I}}\Omega_i$, so tangency conditions on $M$ become
conditions on a product set in $\Rset^n$.

First, we study the properties of contingent cones under the adapted chart. We introduce the notion of \emph{geometric derivability} for the elements in the cotingent cone. 
\begin{definition}[Geometric derivability (Definition 6.1 \cite{Rockafellar1998Variational})]
  \label{def:derivable}
Let $K\subseteq\mbb{R}^n$ and $\bar z\in K$. A vector $w\in T_K(\bar z)$ is
\emph{derivable} if there exist $\varepsilon>0$ and $\xi:[0,\varepsilon]\to K$ with
$\xi(0)=\bar z$ and $\lim_{t\downarrow0}\left(\xi(t)-\bar z\right)/t=w$. The set $K$ is
\emph{geometrically derivable} at $\bar z$ if every $w\in T_K(\bar z)$ is derivable.
\end{definition}

In the above definition, no continuity of $\xi$ on $(0,\varepsilon]$ is required. Clarke regular sets, in particular convex sets, are geometrically derivable \cite[Ch.~6]{Rockafellar1998Variational}; so are finite unions of closed convex sets, because the contingent cone of a finite union of closed sets at $\bar z$ is the union of the contingent cones of the members that contain $\bar z$. The following lemma states the regularity property used to decompose the tangent cone of a product set.
\begin{lemma}[Tangent cones of product sets]
\label{lem:product-tangent-cone}
Let $K=K_1\times\cdots\times K_N$ with $K_i\subseteq\mathbb{R}^{n_i}$, and let
$z=(z_1,\dots,z_N)\in K$. Then
\begin{equation}
T_K(z)\subseteq T_{K_1}(z_1)\times\cdots\times T_{K_N}(z_N)\subseteq\operatorname{co}T_K(z).
\label{eq:product-cone-sandwich}
\end{equation}
If, moreover, $K_i$ is geometrically derivable at $z_i$ for every $i\in\mc{I}$ except at
most one, then
\begin{equation}
T_K(z)=T_{K_1}(z_1)\times\cdots\times T_{K_N}(z_N).
\label{eq:product-tangent-cone}
\end{equation}
\end{lemma}
\begin{proof}
If $v_i\in T_{K_i}(z_i)$, there are $h_k\downarrow0$ and $z_i^k\in K_i$ with
$(z_i^k-z_i)/h_k\to v_i$; since $z_j\in K_j$ for $j\neq i$, the points
$z+\jmath_i(z_i^k-z_i)$ lie in $K$, so $\jmath_iv_i\in T_K(z)$. If $v\in T_K(z)$, there are
$h_k\downarrow0$ and $z^k\in K$ with $(z^k-z)/h_k\to v$, and the $i$-th block gives
$(z_i^k-z_i)/h_k\to v_i$ with $z_i^k\in K_i$; this is the first inclusion in
\eqref{eq:product-cone-sandwich}. For the second, $v=\sum_{i\in\mc{I}}\jmath_iv_i$ is a
finite sum of elements of $T_K(z)$, hence lies in $\operatorname{co}T_K(z)$, because the
convex hull of a cone is a convex cone.

For \eqref{eq:product-tangent-cone}, we show $\supseteq$. Let $i_0\in\mc{I}$ be
such that $K_i$ is geometrically derivable at $z_i$ for all $i\neq i_0$, and let
$v\in\prod_{i\in\mc{I}}T_{K_i}(z_i)$. Choose $h_k\downarrow0$ and $z_{i_0}^k\in K_{i_0}$ with
$(z_{i_0}^k-z_{i_0})/h_k\to v_{i_0}$. For $i\neq i_0$, let $\xi_i:[0,\varepsilon_i]\to K_i$ be
as in Definition~\ref{def:derivable} for $v_i$, and set $z_i^k:=\xi_i(h_k)$ for all $k$
large enough that $h_k\leq\varepsilon_i$. Then $z^k:=(z_1^k,\dots,z_N^k)\in K$ and
$(z^k-z)/h_k\to v$, so $v\in T_K(z)$.
\end{proof}

\subsection{Main contribution}
For $x\in U$, the subsystem velocities in adapted coordinates are
\begin{equation}
\widehat F_i\left(x,u\right)
:=d\pi_i|_{\Phi(x)}\left(d\Phi_xF\left(x,u\right)\right)
\in T_{z_i}V_i,
\label{eq:projected-velocity}
\end{equation}
so that $d\Phi_xF(x,u)=(\widehat F_1(x,u),\dots,\widehat F_N(x,u))$ and $\dot
z_i=\widehat F_i(x,u)$ along trajectories in $U$. By Lemma~\ref{lem:coordinate-projections}, $\widehat F_i\left(x,u\right)$ is the $i$-th coordinate block of the projected velocity $d\Phi_xF_i\left(x,u\right)$.
\begin{theorem}[Control invariance of a local product set]
\label{thm:local-product-invariance}
Let $\Phi=(z_1,\dots,z_N):U\to V_1\times\cdots\times V_N$ be an adapted product chart and let
$\Omega$ be as in \eqref{eq:candidate-set}. Suppose that for every $x\in\Omega$ there exists
$u\in\mbb{U}$ such that
\begin{equation}
\widehat F_i(x,u)\in T_{\Omega_i}\bigl(z_i(x)\bigr)\quad\text{for all } i\in\mc{I}.
\label{eq:projected-tangency}
\end{equation}
Then:
\begin{enumerate}[i)]
\item $F(x,u)\in\operatorname{co}T_\Omega(x)$ for the same $u$; if, in addition, every
$\Omega_i$ except at most one is geometrically derivable at $z_i(x)$, then
$F(x,u)\in T_\Omega(x)$;
\item $\Omega$ is control invariant.
\end{enumerate}
\end{theorem}

\begin{proof}
  $i)$ Fix $x\in\Omega$ and a corresponding $u$, and write $z=\Phi(x)$. By
\eqref{eq:projected-velocity} and \eqref{eq:projected-tangency},
$d\Phi_xF(x,u)\in\prod_{i\in\mc{I}}T_{\Omega_i}(z_i)$, which is contained in
$\operatorname{co}T_{\Phi(\Omega)}(z)$ by Lemma~\ref{lem:product-tangent-cone} and equals
$T_{\Phi(\Omega)}(z)$ under the derivability assumption. Since $d\Phi_x$ is a linear
isomorphism, \eqref{eq:cone-transformation} gives
$d\Phi_x^{-1}\operatorname{co}T_{\Phi(\Omega)}(z)=\operatorname{co}T_\Omega(x)$, and $i)$
follows. $ii)$ By $i)$, Condition~\eqref{eq:controlled-nagumo} holds on $\Omega$. Since $\Omega$ is compact and contained in the domain of the chart $\Phi$, Lemma~\ref{lem:viability} applies with $(O,\varphi)=(U,\Phi)$.
\end{proof}
The input in \eqref{eq:projected-tangency} is common to all subsystems: feasibility of each condition with a different input does not imply \eqref{eq:projected-tangency}. Theorem~\ref{thm:local-product-invariance} is stated in the chart. We now describe $\Omega$ and condition \eqref{eq:projected-tangency} on $M$. For $x\in\Omega$, let
$A(x)=\{i\in\mc{I}\colon z_i(x)\in\partial\Omega_i\}$ be the set of \emph{active} subsystems, and let $\Sigma_i=\{x\in\Omega\colon z_i(x)\in\partial\Omega_i\}$ be the $i$-th \emph{face} of $\Omega$.

\begin{proposition}[The candidate set on $M$]
\label{prop:omega-on-M}
Let $\Phi=(z_1,\dots,z_N)$ be an adapted product chart and let $\Omega$ be as in
\eqref{eq:candidate-set}. Then:
\begin{enumerate}[i)]
\item $\Omega=\bigcap_{i\in\mc{I}}z_i^{-1}(\Omega_i)$, and every level set of $z_i$ is an
integral manifold of $\mc{D}_i^c$. The set $\partial\Omega$ is $\bigcup_{i\in\mc{I}}\Sigma_i$,
and each face $\Sigma_i$ is a union of pieces of integral manifolds of $\mc{D}_i^c$;
\item for $x\in\Omega$ and $v\in\TxM{x}{M}$, $\hat\Pi_i(x)v\in T_\Omega(x)$ iff
$dz_i|_xv\in T_{\Omega_i}(z_i(x))$, and $v\in T_\Omega(x)$ implies
$\hat\Pi_i(x)v\in T_\Omega(x)$ for every $i\in\mc{I}$;
\item if $A(x)=\{i\}$, then
$T_\Omega(x)=\{v\in\TxM{x}{M}:\hat\Pi_i(x)v\in T_\Omega(x)\}\supseteq\mc{D}_i^c(x)$; if
$A(x)=\varnothing$, then $T_\Omega(x)=\TxM{x}{M}$.
\end{enumerate}
\end{proposition}
\begin{proof}
$i)$ Since $\Omega\subset U$, $\Phi(\Omega)=\prod_i\Omega_i$ is equivalent to
$\Omega=\bigcap_iz_i^{-1}(\Omega_i)$. Each $z_i$ is a submersion with $\ker dz_i=\mc{D}_i^c$, so its level sets are tangent to $\mc{D}_i^c$. Since $\Phi$ is a homeomorphism onto the open set $V_1\times\cdots\times V_N$ and each $\Omega_i$ is closed, $\partial\Omega=\Phi^{-1}\bigl(\bigcup_i\partial\Omega_i\times\prod_{j\neq i}\Omega_j\bigr)
=\bigcup_i\Sigma_i$, and $\Sigma_i=\bigcup_{c\in\partial\Omega_i}\bigl(\Omega\cap
z_i^{-1}(c)\bigr)$.
For $ii)$, Let $z=\Phi(x)$. By Lemma~\ref{lem:coordinate-projections},
$d\Phi_x\hat\Pi_i(x)v=\jmath_i\,dz_i|_xv$. By Lemma~\ref{lem:product-tangent-cone},
$\jmath_iw\in T_{\Phi(\Omega)}(z)$ iff $w\in T_{\Omega_i}(z_i)$: sufficiency is
$\jmath_iT_{\Omega_i}(z_i)\subseteq T_{\Phi(\Omega)}(z)$, and necessity is the first inclusion
in \eqref{eq:product-cone-sandwich}. With \eqref{eq:cone-transformation}, this gives the
equivalence. If $v\in T_\Omega(x)$, then $d\Phi_xv\in T_{\Phi(\Omega)}(z)\subseteq
\prod_jT_{\Omega_j}(z_j)$, so $dz_i|_xv\in T_{\Omega_i}(z_i)$ and $\hat\Pi_i(x)v\in
T_\Omega(x)$.

Lastly, for $iii)$, $j\notin A(x)$, $z_j$ is an interior point of $\Omega_j$, so
$T_{\Omega_j}(z_j)=\Rset^{n_j}$ and $\Omega_j$ is geometrically derivable at $z_j$. If
$A(x)=\{i\}$, \eqref{eq:product-tangent-cone} and \eqref{eq:cone-transformation} give
$T_\Omega(x)=\{v:dz_i|_xv\in T_{\Omega_i}(z_i)\}$, which equals
$\{v:\hat\Pi_i(x)v\in T_\Omega(x)\}$ by (b) and contains $\ker dz_i|_x=\mc{D}_i^c(x)$. If
$A(x)=\varnothing$, then $x$ is an interior point of $\Omega$.
\end{proof}

\subsection{Blockwise control feasibility}
To obtain a blockwise control structure compatible with the tangent bundle splitting, suppose that
\begin{equation}
\mbb{U}=\mbb{U}_1\times\cdots\times\mbb{U}_N,\qquad g(x)u=\sum_{i\in\mc{I}}G_i(x)u_i,
\label{eq:blockwise-inputs}
\end{equation}
where $u=(u_1,\dots,u_N)$, each $\mbb{U}_i\subset\Rset^{m_i}$ is nonempty, convex and
compact, $\sum_{i\in\mc{I}}m_i=m$, and each $G_i(x):\Rset^{m_i}\to\TxM{x}{M}$ satisfies
\begin{equation}
\operatorname{Im}G_i(x)\subseteq\mc{D}_i(x),\qquad x\in U.
\label{eq:input-alignment}
\end{equation}
Equivalently, $\Pi_j(x)G_i(x)=0$ for $j\neq i$. Then $dz_i|_xG_j(x)=0$ for $j\neq i$, so
$\widehat F_i$ depends on the input only through $u_i$, and we write
\begin{equation}
\widehat F_i(x,u_i):=dz_i|_x\bigl(f(x)+G_i(x)u_i\bigr).
\label{eq:blockwise-velocity}
\end{equation}
\begin{corollary}[Blockwise feasibility]
\label{cor:blockwise-feasibility}
Let $\Phi$ and $\Omega$ be as in Theorem~\ref{thm:local-product-invariance}, and assume
\eqref{eq:blockwise-inputs} and \eqref{eq:input-alignment}. For $x\in\Omega$, let
\begin{equation*}
\mc{K}_i(x):=\bigl\{u_i\in\mbb{U}_i:\widehat F_i(x,u_i)\in T_{\Omega_i}\bigl(z_i(x)\bigr)\bigr\}.
\end{equation*}
If $\mc{K}_i(x)\neq\varnothing$ for every $x\in\Omega$ and every $i\in A(x)$, then $\Omega$ is control invariant.
\end{corollary}
\begin{proof}
Fix $x\in\Omega$. Choose $u_i\in\mc{K}_i(x)$ for $i\in A(x)$, and $u_i\in\mbb{U}_i$
arbitrary otherwise. Then $u:=(u_1,\dots,u_N)\in\mbb{U}$ by \eqref{eq:blockwise-inputs}, and $\widehat F_i(x,u)=\widehat F_i(x,u_i)$ by \eqref{eq:blockwise-velocity}, so the $i$-th condition in \eqref{eq:projected-tangency} holds for $i\in A(x)$; otherwise, it holds because $T_{\Omega_i}(z_i(x))=\Rset^{n_i}$. Theorem~\ref{thm:local-product-invariance}
applies.
\end{proof}

\begin{remark}[Decentralised feedback]
Corollary~\ref{cor:blockwise-feasibility} reduces the viability condition \eqref{eq:controlled-nagumo} to separate input choices when state constraints have an adapted product form, input constraints are independent, and \eqref{eq:input-alignment} holds. This separates control channels without requiring independent subsystem dynamics or ensuring decentralised feedback, since $\mc{K}_i(x)$ may depend on the full state. Writing
$z=(z_i,z_{-i})$, a decentralised selection $u_i=\kappa_i(z_i)$ must satisfy $\widehat F_i\bigl(\Phi^{-1}(z_i,z_{-i}),\kappa_i(z_i)\bigr)\in T_{\Omega_i}(z_i)$ for all $z_i\in\partial\Omega_i,\ z_{-i}\in\textstyle\prod_{j\neq i}\Omega_j$. Theorem~\ref{thm:local-product-invariance} guarantees admissible input trajectories. Additional regularity conditions are needed to ensure that a state-dependent selection yields well-posed closed-loop dynamics.
\end{remark}

\section{Illustrative example}
\label{sec:examples}
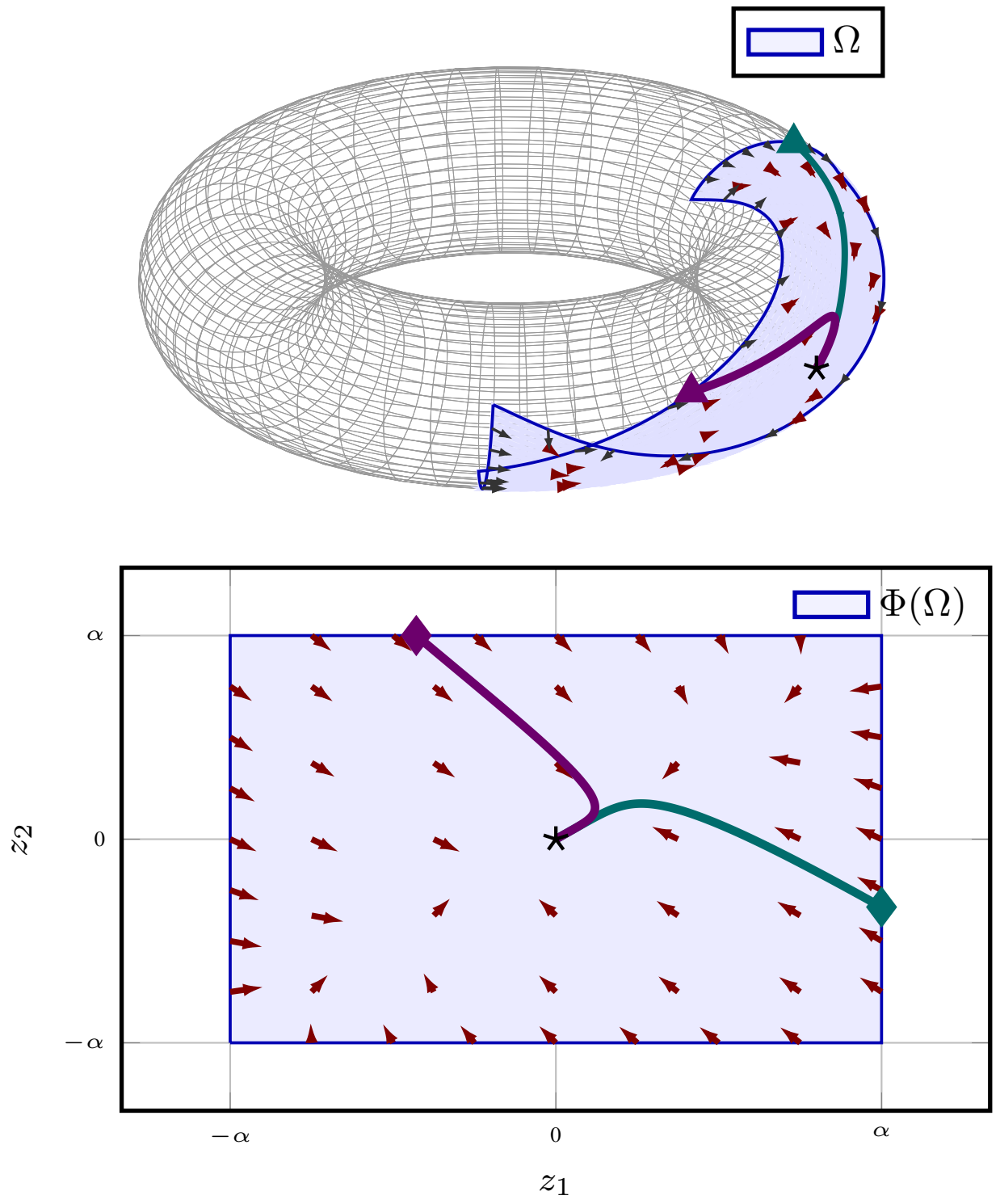
\begin{figure}[t!]
  \centering
  \input{E2}
  \caption{Adapted coordinates for the Kuramoto oscillator in \eqref{eq:example-phase-dynamics}. (top) Invariant set $\Omega$ on $\mbb{T}^2$ with two trajectories \eqref{fig:ex:theta_1} and \eqref{fig:ex:theta_2} starting from the initial conditions $z^{(1)}\in\Phi(\Omega)$ \eqref{fig:ex:theta_1_0}  and $z^{(2)} \in\Phi(\Omega)$\eqref{fig:ex:theta_2_0} respectively towards $z^\ts{eq}$ \eqref{fig:ex:theta_eq}. (bottom) Adapted chart image $\Phi(\Omega)$.  For both $\Omega$ and $\Phi(\Omega)$, the vector field \eqref{fig:ex:vf_torus} shows that all flows remain within these sets.
  }
  \label{fig:example-2}
\end{figure}

Consider the following Kuramoto oscillator comprising of two interconnected subsystems
\begin{subequations}
\begin{align}
\dot\theta_1
&=a_1+k\sin(\theta_2-\theta_1)+u_1,\\
\dot\theta_2
&=a_2+k\sin(\theta_1-\theta_2)+u_2,
\end{align}
\label{eq:example-phase-dynamics}  
\end{subequations}
defined on $\mbb{T}^2=\mbb{S}^1\times\mbb{S}^1$ subject to input constraints $u_i\in\mbb{U}_i=[-\bar u_i,\bar u_i]$, with $\bar u_i>0$ and $i=1,2$. The parameters of the system are: $a_1,a_2\in\Rset$ are the drift and $k\geq0$ the coupling strength. The distributions $\mc{D}_i=\ts{span}\{\partial_{\theta_i}\}$, $i\in\{1,2\}$, satisfy $\TxM{}{\mbb{T}^2} =\mc{D}_1\oplus\mc{D}_2,$ and $[\partial_{\theta_1},\partial_{\theta_2}]=0$. The adapted chart $\Phi:U\to(-\pi,\pi)^2$, mapping phase arcs to intervals, assigns each phase a unique $z_i \in \left(-\pi, \pi\right)$, see Figure~\ref{fig:example-2} top and middle. We, now, construct the adapted chart by defining the angular variables $z_1 = \theta_1$, and $z_2=\theta_2-\beta\sin(\theta_1)$ modulo $2\pi$ and  $\beta>0$. This transformation is a smooth diffeomorphism on $\mbb{T}^2$ with inverse $\theta_1=z_1$, $\theta_2=z_2+\beta\sin(z_1)$ modulo $2\pi$. The associated vector fields and distributions are: $X_1=\partial_{\theta_1}+\beta\cos\theta_1\,\partial_{\theta_2}$, $X_2=\partial_{\theta_2}$ and let $\mathcal{D}_1=\ts{span}\{X_1\}$ and $\mc{D}_2=\ts{span}\{X_2\}$. These satisfy $\TxM{}{\mbb{T}^2} = \mc{D}_1\oplus\mc{D}_2$ following linear independence of $X_1$ and $X_2$; additionally, we have $[X_1,X_2]=0$ and $d\Phi_x\left (X_1(\theta)\right )=(1,0)$ and $d\Phi_x\left(X_2(\theta)\right)=(0,1)$. Therefore, $\Phi$ is adapted to the splitting because its differential maps each distribution $\mathcal{D}_i\left(x\right)$ onto the corresponding coordinate direction in the $z$ coordinates. The new dynamics are:
\begin{subequations}
  \begin{align}
\dot z_1
&=b_1(z) + u_1,\\
\dot z_2
&=\beta\cos(z_1)(b_1(z)+u_1) + b_2(z)+u_2,
  \end{align}
  \label{eq:transformed-dynamics}
\end{subequations}
where $b_1(z) =a_1+k\sin(z_2-\beta\sin(z_1)-z_1)$, $b_2(z)=a_2+k\sin(z_1-z_2+\beta\sin(z_1))$, and $u_i\in[-\bar u_i,\bar u_i]$ and $k\geq0$. Under this transformation, each input map, however, is aligned with its corresponding distribution, \ie  $u_1$ to $\mc{D}_1$ and $u_2$ to $\mc{D}_2$ respectively.

The constraints on $\mbb{T}^2$ for $\alpha \in (0,\tfrac{\pi}{2})$ are $|\theta_1|\leq \alpha$ and $|\theta_2-\beta\sin\theta_1|\leq\alpha$. The set $\Omega$ is clearly nonconvex, but the adapted charts transform it into a square. The inputs $u_i=-a_i$ satisfy the tangent conditions whenever $|a_i|\leq\bar u_i$. Since the input maps are aligned with the splitting, Corollary~\ref{cor:blockwise-feasibility} provides control invariance for the entire set. We note that sets $[-\alpha,\alpha]\times[c,d]\subset\mbb{T}^2$ with $c \leq d$ cannot be made invariant. 
For invariance to hold, $[c,d]$ must be contained in every interval $[\beta\sin\left(\theta_1\right)-\alpha,\,\beta\sin\left(\theta_1\right)+\alpha]$, $\theta_1\in[-\alpha,\alpha]$. Since $\sin\left(\theta_1\right)$ is increasing on this interval, a nonempty intersection requires $\beta\sin\left(\alpha\right)-\alpha\leq\alpha-\beta\sin\left(\alpha\right)$. Due to this, if $\beta\sin(\alpha)>\alpha$, then no nonempty rectangle contained in $\Omega$ can keep the full range $[-\alpha,\alpha]$ of the first phase. 

The example shows how adapted coordinates enable subsystem tangency conditions to certify the entire curved region without restricting it to a rectangle in original coordinates. The synchronisation conditions in \cite{Doerfler2012} do not directly certify this region, while the invariant set construction presented in \cite{Nilsson2016} only applies to coupled linear systems.

\section{Conclusion and future work}
\label{sec:conclusion}
This paper addressed the issue of design of safety sets from a distributed point of view. We generalised the Euclidean notion of coordinates, \ie assignment of vector field coordinates to agents, using distributions of the tangent bundle. Using these distributions, we have constructed control invariant sets with a product structure through a tangent bundle splitting and adapted coordinates. The torus example shows how adapted coordinates represent an entire curved constraint region as a product, whereas no contained rectangle in the original coordinates spans the full first phase range under the stated conditions. The construction, therefore, preserves the full constraint region while allowing for subsystem boundary checks. Future work will address the selection of compatible distributions and constraint sets, and extension to larger regions with complex or learned geometry.

\bibliography{biblio}
\end{document}

%% file: adaptedproductchart.tex
    \begin{tikzpicture}[
    >=Latex,
    every node/.style={font=\small},
      leafA/.style={dOneLight, line width=0.35pt},
      leafB/.style={dTwoLight, line width=0.35pt},
      vec/.style={->, line width=0.9pt},
      lbl/.style={font=\footnotesize}
    ]
\def\PX#1#2{(#1)+0.2*(#2)*(#2)}
\def\PY#1#2{(#2)-0.15*(#1)*(#1)}
\def\sa{-0.9}\def\sb{0.9}\def\ta{-0.8}\def\tb{1.0}

\colorlet{dOne}{blue!85!black}
\colorlet{dTwo}{teal!85!black}
\colorlet{dOneLight}{blue!22}
\colorlet{dTwoLight}{teal!32}
\colorlet{faceCol}{teal!45}

\begin{scope}[scale=0.9, every node/.append style={transform shape}]

\begin{scope}[shift={(-1,0)},scale = 1.1]
  \draw[black!45, fill=black!2] plot[smooth cycle, tension=0.75] coordinates
    {(-1.95,-1.95) (-2.05,0.95) (-1.1,1.95) (0.9,1.85) (2.35,1.05)
     (2.45,-1.1) (1.3,-2.3) (-0.7,-2.35)};
  \node[lbl, black!55] at (-2,1.7) {$M$};
  \fill[white] plot[domain=-1.5:1.5, samples=30, variable=\s] ({\PX{\s}{-1.5}},{\PY{\s}{-1.5}})
    -- plot[domain=-1.5:1.5, samples=30, variable=\t] ({\PX{1.5}{\t}},{\PY{1.5}{\t}})
    -- plot[domain=1.5:-1.5, samples=30, variable=\s] ({\PX{\s}{1.5}},{\PY{\s}{1.5}})
    -- plot[domain=1.5:-1.5, samples=30, variable=\t] ({\PX{-1.5}{\t}},{\PY{-1.5}{\t}}) -- cycle;
  \fill[blue!8] plot[domain=\sa:\sb, samples=25, variable=\s] ({\PX{\s}{\ta}},{\PY{\s}{\ta}})
    -- plot[domain=\ta:\tb, samples=25, variable=\t] ({\PX{\sb}{\t}},{\PY{\sb}{\t}})
    -- plot[domain=\sb:\sa, samples=25, variable=\s] ({\PX{\s}{\tb}},{\PY{\s}{\tb}})
    -- plot[domain=\tb:\ta, samples=25, variable=\t] ({\PX{\sa}{\t}},{\PY{\sa}{\t}}) -- cycle;
  \foreach \c in {-1.2,-0.6,0,0.6,1.2} {
    \draw[leafA] plot[domain=-1.5:1.5, samples=30, variable=\s] ({\PX{\s}{\c}},{\PY{\s}{\c}});
    \draw[leafB] plot[domain=-1.5:1.5, samples=30, variable=\t] ({\PX{\c}{\t}},{\PY{\c}{\t}});
  }
  \draw[blue, line width=0.6pt] plot[domain=\sb:\sa, samples=25, variable=\s] ({\PX{\s}{\tb}},{\PY{\s}{\tb}})
    -- plot[domain=\tb:\ta, samples=25, variable=\t] ({\PX{\sa}{\t}},{\PY{\sa}{\t}})
    -- plot[domain=\sa:\sb, samples=25, variable=\s] ({\PX{\s}{\ta}},{\PY{\s}{\ta}});
  \draw[faceCol, line width=2.4pt] plot[domain=\ta:\tb, samples=25, variable=\t] ({\PX{\sb}{\t}},{\PY{\sb}{\t}});
  \draw[black, line width=0.7pt] plot[domain=-1.5:1.5, samples=30, variable=\s] ({\PX{\s}{-1.5}},{\PY{\s}{-1.5}})
    -- plot[domain=-1.5:1.5, samples=30, variable=\t] ({\PX{1.5}{\t}},{\PY{1.5}{\t}})
    -- plot[domain=1.5:-1.5, samples=30, variable=\s] ({\PX{\s}{1.5}},{\PY{\s}{1.5}})
    -- plot[domain=1.5:-1.5, samples=30, variable=\t] ({\PX{-1.5}{\t}},{\PY{-1.5}{\t}}) -- cycle;
  \node[lbl] at (-1.3,1.25) {$U$};
  \node[lbl] at ({\PX{-0.3}{0.65}},{\PY{-0.1}{0.55}}) {$\Omega$};
  \node[lbl, dTwo] at (1.33,-0.55) {$\Sigma_1$};
  \coordinate (x) at ({\PX{-0.45}{-0.4}},{\PY{-0.5}{-0.4}});
  \draw[vec, dOne] (x) -- ++(0.445,0.067) node[right, lbl, inner sep=1pt] {$\mc{D}_1(x)$};
  \draw[vec, dTwo] (x) -- ++(-0.071,0.444) node[above, lbl, inner sep=1pt] {$\mc{D}_2(x)$};
  \fill (x) circle (1.3pt) node[below left, lbl, inner sep=1pt] {$x$};
  \coordinate (p) at ({\PX{\sb}{0.5}},{\PY{\sb}{0.5}});
  \draw[vec, dOne] (p) -- ++(-0.483,0.130) node[left, lbl, inner sep=1pt] {$F_1$};
  \draw[vec, black] (p) -- ++(-0.408,0.503) node[left, lbl, inner sep=1pt] {$F$};
  \draw[vec, dTwo] (p) -- ++(0.0745,0.373) node[right, lbl, inner sep=1pt] {$F_2$};
  \fill (p) circle (1.3pt);
    \end{scope}

\draw[->, line width=1pt] (2.75,0.45) -- node[above, lbl] {$\Phi$} (4.15,0.45);
\draw[->, line width=1pt] (4.15,-0.45) -- node[below, lbl] {$\Phi^{-1}$} (2.75,-0.45);

\begin{scope}[shift={(7.55,0)},scale = 1.25]
  \fill[white] (-1.5,-1.5) rectangle (1.5,1.5);
  \fill[blue!8] (\sa,\ta) rectangle (\sb,\tb);
  \foreach \c in {-1.2,-0.6,0,0.6,1.2} {
    \draw[leafA] (-1.5,\c) -- (1.5,\c);
    \draw[leafB] (\c,-1.5) -- (\c,1.5);
  }
  \draw[black, line width=0.6pt] (\sb,\tb) -- (\sa,\tb) -- (\sa,\ta) -- (\sb,\ta);
  \draw[faceCol, line width=2.4pt] (\sb,\ta) -- (\sb,\tb);
  \draw[black, line width=0.7pt] (-1.5,-1.5) rectangle (1.5,1.5);
  \node[lbl] at (0,1.8) {$V_1\times V_2$};
  \node[lbl] at (-0.38,0.75) {$\Omega_1\times\Omega_2$};
  \draw[line width=1.4pt] (\sa,-1.75) -- (\sb,-1.75) node[midway, below, lbl] {$\Omega_1$};
  \draw[line width=1.4pt] (-1.75,\ta) -- (-1.75,\tb) node[midway, left, lbl] {$\Omega_2$};
  \node[lbl, black!60] at (1.75,-1.75) {$z_1$};
  \node[lbl, black!60] at (-1.75,1.75) {$z_2$};
  \node[lbl, dTwo, rotate=90] at (1.2,-0.35) {$\partial\Omega_1\times\Omega_2$};
  \coordinate (z) at (-0.5,-0.4);
  \draw[vec, dOne] (z) -- ++(0.45,0) node[right, lbl, inner sep=1pt] {$E_1$};
  \draw[vec, dTwo] (z) -- ++(0,0.45) node[above right, lbl, inner sep=1pt] {$E_2$};
  \fill (z) circle (1.3pt) node[below left, lbl, inner sep=1pt] {$z$};
  \coordinate (q) at (\sb,0.5);
  \draw[vec, dOne] (q) -- ++(-0.5,0) node[left, lbl, inner sep=1pt] {$\widehat F_1$};
  \draw[vec, black] (q) -- ++(-0.5,0.38);
  \draw[vec, dTwo] (q) -- ++(0,0.38) node[right, lbl, inner sep=1pt] {$\widehat F_2$};
  \fill (q) circle (1.3pt);
\end{scope}

\end{scope}

    \end{tikzpicture}


%% file: E2.tex
    \begin{tikzpicture}[
        >=Latex,
        every node/.style={font=\small}
          ]
    \begin{axis}[
        name=torusfield,
        at={(0,0cm)},
        width=0.45\textwidth,
        height=0.35\textwidth,
        view={35}{25},
        axis equal image,
        hide axis,
        xmin=-3,
        xmax=3,
        ymin=-3,
        ymax=3,
        zmin=-1.4,
        zmax=1.4,
        clip=false,
        colormap={safepatch}{
            color(0cm)=(blue!12);
            color(1cm)=(blue!12)
        }
        ]
        \addplot3[
            mesh,
            mesh/rows=49,
            mesh/cols=49,
            draw=black!40,
            line width=0.15pt,
            forget plot
        ]
            table[x=x,y=y,z=z]
            {torus_mesh.dat};
        \addplot3[
            surf,
            shader=flat,
            mesh/rows=31,
            mesh/cols=31,
            draw=none,
            fill opacity=0.65,
            z buffer=sort,
            forget plot
        ]
            table[x=x,y=y,z=z]
            {safe_patch.dat};

        \addplot3[
            blue!70!black,
            line width=0.7pt,
            forget plot
        ]
            table[x=x,y=y,z=z]
            {torus_boundary.dat};
            
        \addplot3[
            red!50!black,
            line width=1.4pt,
            -{Latex[length=1.9mm,width=1mm]},
            quiver={
                u=\thisrow{dx},
                v=\thisrow{dy},
                w=\thisrow{dz},
                scale arrows=1
            },
            forget plot
        ]
            table[x=x,y=y,z=z]
            {interior_field.dat};
            \label{fig:ex:vf_torus}

        \addplot3[
            black!80,
            line width=0.65pt,
            -{Latex[length=1.2mm,width=0.75mm]},
            quiver={
                u=\thisrow{dx},
                v=\thisrow{dy},
                w=\thisrow{dz},
                scale arrows=1.5
            },
            forget plot
        ]
            table[x=x,y=y,z=z]
            {boundary_field.dat};                                                                          

        \addplot3[
            teal!85!black,
            trajectory path,
            line width=1.5pt,
            forget plot
        ]
            table[x=x,y=y,z=z]
            {torus_trajectory_1.dat};
            \label{fig:ex:theta_1}

        \addplot3[
            violet!85!black,
            trajectory path, 
            line width=2pt,
            forget plot
        ]
            table[x=x,y=y,z=z]
            {torus_trajectory_2.dat};
            \label{fig:ex:theta_2}

        \addplot3[
            teal!85!black,
            only marks,
            mark=triangle*,
            mark size=4pt,
            thick,
            forget plot
        ]
            table[x=x,y=y,z=z]
            {torus_start_1.dat};

        \addplot3[
            violet!85!black,
            only marks,
            mark=triangle*,
            mark size=4pt,
            thick,
            forget plot
        ]
            table[x=x,y=y,z=z]
            {torus_start_2.dat};

        \addplot3[
            black,
            only marks,
            mark=star,
            mark size=3pt,
            thick,
            forget plot
        ]
            table[x=x,y=y,z=z]
            {equilibrium_point.dat};

            \addlegendimage{
              fill=blue!8,
              fill opacity = 0.65,
              draw=blue!70!black,
              area legend
            }
            \addlegendentry{$\Omega$}

          \end{axis}
          
    \begin{axis}[
        name=adaptedfield,
        at={(0cm,-4.5cm)},
        width=0.4\textwidth,
        height=0.25\textwidth,
        scale only axis,
        xmin=-1.4,
        xmax=1.4,
        ymin=-1.4,
        ymax=1.4,
        grid=major,
        xlabel={$z_1$},
        ylabel={$z_2$},
        xtick={-1.05,0,1.05},
        xticklabels={$-\alpha$,0,$\alpha$},
        ytick={-1.05,0,1.05},
        yticklabels={$-\alpha$,0,$\alpha$},
        axis lines=box,
        clip=false,
        legend style={
          at={(0.878,1)},
          anchor=north,
          draw=none,
          fill = none,
          font=\tiny,
          /tikz/every even column/.append style={column sep=8pt}
        }
        ]

        \addplot[
            draw=blue!70!black,
            fill=blue!8,
            line width=0.7pt,
            forget plot
        ]
            table[x=z1,y=z2]
            {adapted_square.dat};
        \addplot[
            red!50!black,
            line width=1.4pt,
            -{Latex[length=1.9mm,width=1mm]},
            quiver={
                u=\thisrow{dz1},
                v=\thisrow{dz2},
                scale arrows=1
            },
            forget plot
        ]
            table[x=z1,y=z2]
            {adapted_interior_field.dat};
        \addplot[
            red!50!black,
            line width=1.4pt,
            -{Latex[length=1.9mm,width=1mm]},
            quiver={
                u=\thisrow{dz1},
                v=\thisrow{dz2},
                scale arrows=1
            },
            forget plot
        ]
            table[x=z1,y=z2]
            {adapted_boundary_field.dat};

        \addplot[
            teal!85!black,
            trajectory path,
            line width=2pt,
            forget plot
        ]
            table[x=z1,y=z2]
            {adapted_trajectory_1.dat};

        \addplot[
            violet!85!black,
            trajectory path, 
            line width=2pt,
            forget plot
        ]
            table[x=z1,y=z2]
            {adapted_trajectory_2.dat};

        \addplot[
            teal!85!black,
            only marks,
            mark=diamond*,
            mark size=4pt,
            mark options={fill=teal!85!black},
            forget plot
        ]
            table[x=z1,y=z2]
            {adapted_start_1.dat};
            \label{fig:ex:theta_1_0}
        \addplot[
        violet!85!black,
        mark options={fill=violet!85!black},
            only marks,
            mark=diamond*,
            mark size=4pt,
            forget plot
        ]
            table[x=z1,y=z2]
            {adapted_start_2.dat};
            \label{fig:ex:theta_2_0}
        \addplot[
            black,
            only marks,
            mark=star,
            mark size=3pt,
            thick,
            forget plot
        ]
            table[x=z1,y=z2]
            {adapted_equilibrium.dat};
            \label{fig:ex:theta_eq}
            
            \addlegendimage{
             fill=blue!8,
              fill opacity = 0.65,
              draw=blue!70!black,
              area legend
            }
            \addlegendentry{$\Phi(\Omega)$}
      \end{axis}
      
    \end{tikzpicture}


    

%% file: main.bbl
\begin{thebibliography}{10}
\providecommand{\url}[1]{#1}
\csname url@samestyle\endcsname
\providecommand{\newblock}{\relax}
\providecommand{\bibinfo}[2]{#2}
\providecommand{\BIBentrySTDinterwordspacing}{\spaceskip=0pt\relax}
\providecommand{\BIBentryALTinterwordstretchfactor}{4}
\providecommand{\BIBentryALTinterwordspacing}{\spaceskip=\fontdimen2\font plus
\BIBentryALTinterwordstretchfactor\fontdimen3\font minus \fontdimen4\font\relax}
\providecommand{\BIBforeignlanguage}[2]{{%
\expandafter\ifx\csname l@#1\endcsname\relax
\typeout{** WARNING: IEEEtran.bst: No hyphenation pattern has been}%
\typeout{** loaded for the language `#1'. Using the pattern for}%
\typeout{** the default language instead.}%
\else
\language=\csname l@#1\endcsname
\fi
#2}}
\providecommand{\BIBdecl}{\relax}
\BIBdecl

\bibitem{Blanchini1999}
F.~Blanchini, ``Set invariance in control,'' \emph{Automatica}, vol.~35, no.~11, pp. 1747--1767, 1999.

\bibitem{aubin2009viability}
J.-P. Aubin, \emph{Viability Theory}, ser. Modern Birkh{\"a}user Classics.\hskip 1em plus 0.5em minus 0.4em\relax Boston: Birkh{\"a}user, 2009.

\bibitem{Hernandez2016}
B.~Hernandez, P.~Baldivieso-Monasterios, and P.~Trodden, ``{Distributed MPC: Guaranteeing Global Stability from Locally Designed Tubes},'' \emph{IFAC-PapersOnLine}, vol.~50, no.~1, pp. 11\,829--11\,834, jul 2017.

\bibitem{Trodden2026}
P.~Trodden, J.~M. Maestre, and H.~Ishii, ``On the limits of robust control under adversarial disturbances,'' \emph{IEEE Transactions on Automatic Control}, pp. 1--16, 2026.

\bibitem{Baldivieso-Monasterios2021}
P.~R. Baldivieso-Monasterios and P.~A. Trodden, ``{Coalitional predictive control: Consensus-based coalition forming with robust regulation},'' \emph{Automatica}, vol. 125, p. 109380, 2021.

\bibitem{Masero2023}
E.~Masero, P.~R. Baldivieso-Monasterios, J.~M. Maestre, and P.~A. Trodden, ``{Robust coalitional model predictive control with plug-and-play capabilities},'' \emph{Automatica}, vol. 153, p. 111053, 2023.

\bibitem{Ames2016}
A.~D. Ames, X.~Xu, J.~W. Grizzle, and P.~Tabuada, ``Control barrier function based quadratic programs for safety critical systems,'' \emph{IEEE Transactions on Automatic Control}, vol.~62, no.~8, pp. 3861--3876, 2017.

\bibitem{Rabiee2025}
P.~Rabiee and J.~B. Hoagg, ``Soft-minimum and soft-maximum barrier functions for safety with actuation constraints,'' \emph{Automatica}, vol. 171, p. 111921, Jan. 2025.

\bibitem{Prajna2007}
S.~Prajna, A.~Jadbabaie, and G.~J. Pappas, ``A framework for worst-case and stochastic safety verification using barrier certificates,'' \emph{IEEE Transactions on Automatic Control}, vol.~52, no.~8, pp. 1415--1428, 2007.

\bibitem{WANG2025112393}
H.~Wang, A.~Papachristodoulou, and K.~Margellos, ``Distributed safe control design and probabilistic safety verification for multi-agent systems,'' \emph{Automatica}, vol. 179, p. 112393, 2025.

\bibitem{Bansal2018}
S.~Bansal, M.~Chen, S.~Herbert, and C.~J. Tomlin, ``{Hamilton-jacobi reachability: A brief overview and recent advances},'' \emph{2017 IEEE 56th Annu. Conf. Decis. Control. CDC 2017}, vol. 2018-Janua, pp. 2242--2253, 2018.

\bibitem{Nilsson2016}
P.~Nilsson and N.~Ozay, ``Synthesis of separable controlled invariant sets for modular local control design,'' in \emph{2016 American Control Conference (ACC)}.\hskip 1em plus 0.5em minus 0.4em\relax IEEE, 2016, pp. 5656--5663.

\bibitem{Pirastehzad2026}
A.~Pirastehzad, A.~van~der Schaft, and B.~Besselink, ``Modular verification and design of interconnected linear systems using $(\gamma,\delta)$-similarity,'' \emph{Automatica}, vol. 190, p. 113046, Aug. 2026.

\bibitem{Jurdjevic1996}
V.~Jurdjevic, \emph{Geometric Control Theory}.\hskip 1em plus 0.5em minus 0.4em\relax Cambridge University Press, 1996.

\bibitem{Bullo2005}
F.~Bullo and A.~D. Lewis, \emph{Geometric Control of Mechanical Systems: Modeling, Analysis, and Design for Simple Mechanical Control Systems}.\hskip 1em plus 0.5em minus 0.4em\relax Springer, 2005, vol.~49.

\bibitem{deSa2025geometric}
M.~de~Sa, P.~Ong, and A.~D. Ames, ``From bundles to backstepping: Geometric control barrier functions for safety-critical control on manifolds,'' \emph{arXiv preprint arXiv:2510.20202}, 2025.

\bibitem{Ibuki2020}
T.~Ibuki, S.~Wilson, A.~D. Ames, and M.~Egerstedt, ``Distributed collision-free motion coordination on a sphere: A conic control barrier function approach,'' \emph{IEEE Control Systems Letters}, vol.~4, no.~4, pp. 976--981, 2020.

\bibitem{Doerfler2012}
F.~D{\"o}rfler and F.~Bullo, ``Synchronization and transient stability in power networks and nonuniform {Kuramoto} oscillators,'' \emph{SIAM Journal on Control and Optimization}, vol.~50, no.~3, pp. 1616--1642, 2012.

\bibitem{lee2011}
T.~Lee, ``Geometric tracking control of the attitude dynamics of a rigid body on {SO(3)},'' in \emph{Proceedings of the 2011 American Control Conference}.\hskip 1em plus 0.5em minus 0.4em\relax IEEE, 2011, pp. 1200--1205.

\bibitem{Abdelgalil2026}
M.~Abdelgalil, M.~Krstic, and J.~I. Poveda, ``Non-holonomic gradient play: Leafwise nash equilibria, stability, and deception,'' Sep. 2026.

\bibitem{doCarmo1992}
M.~P. do~Carmo, \emph{Riemannian Geometry}, ser. Mathematics: Theory \& Applications.\hskip 1em plus 0.5em minus 0.4em\relax Boston: Birkh{\"a}user, 1992.

\bibitem{Lee2018}
J.~M. Lee, \emph{Introduction to Riemannian Manifolds}, 2nd~ed., ser. Graduate Texts in Mathematics.\hskip 1em plus 0.5em minus 0.4em\relax Cham: Springer, 2018, vol. 176.

\bibitem{Lee2013}
------, \emph{Introduction to Smooth Manifolds}, 2nd~ed., ser. Graduate Texts in Mathematics.\hskip 1em plus 0.5em minus 0.4em\relax New York: Springer, 2013, vol. 218.

\bibitem{Rockafellar1998Variational}
R.~T. Rockafellar and R.~J.-B. Wets, \emph{Variational Analysis}.\hskip 1em plus 0.5em minus 0.4em\relax Springer, 1998.

\end{thebibliography}
